# Real-time, inline quantitative MRI enabled by scanner-integrated machine learning: a proof of principle with NODDI

Samuel Rot[1,2], Iulius Dragonu[3], Christina Triantafyllou[3], Matthew Grech-Sollars[1,4], Anastasia Papadaki[4,5], Laura Mancini[4,5], Stephen Wastling[4,5], Jennifer Steeden[6], John S. Thornton[4,5], Tarek Yousry[4,5], Claudia A. M. Gandini Wheeler-Kingshott[2,7,8], David L. Thomas[4], Daniel C. Alexander[1], Hui Zhang[1]

[1]Hawkes Institute, Department of Computer Science, UCL, London, United Kingdom
[2]NMR Research Unit, Queen Square MS Centre, Department of Neuroinflammation, UCL Queen Square Institute of Neurology, Faculty of Brain Sciences, UCL, London, United Kingdom
[3]Research and Collaborations GBI, Siemens Healthcare Ltd, Camberley, United Kingdom
[4]Lysholm Department of Neuroradiology, National Hospital for Neurology and Neurosurgery, University College London Hospitals NHS Foundation Trust, London, UK
[5]Neuroradiological Academic Unit, Dept of Translational Neuroscience and Stroke, UCL Queen Square Institute of Neurology, UCL, London, UK
[6]Centre for Cardiovascular Imaging, Institute of Cardiovascular Science, UCL, London, UK
[7]Department of Brain & Behavioural Sciences, University of Pavia, Pavia, Italy
[8]Digital Neuroscience Centre, IRCCS Mondino Foundation, Pavia, Italy

## ABSTRACT

**Purpose:** The clinical feasibility and translation of many advanced quantitative MRI (qMRI) techniques are inhibited by their restriction to 'research mode', due to resource-intensive, offline parameter estimation. This work aimed to achieve 'clinical mode' qMRI, by real-time, inline parameter estimation with a trained neural network (NN) fully integrated into a vendor's image reconstruction environment, therefore facilitating and encouraging clinical adoption of advanced qMRI techniques.

**Methods:** The Siemens Image Calculation Environment (ICE) pipeline was customised to deploy trained NNs for advanced diffusion MRI parameter estimation with Open Neural Network Exchange (ONNX) Runtime. Two fully-connected NNs were trained offline with data synthesised with the neurite orientation dispersion and density imaging (NODDI) model, using either conventionally estimated ($NN_{MLE}$) or ground truth ($NN_{GT}$) parameters as training labels. The strategy was demonstrated online in two healthy volunteers (one rescanned) and evaluated offline with synthetic data, testing two diffusion protocols.

**Results:** NNs were successfully integrated and deployed natively in ICE, performing inline, whole-brain, in vivo NODDI parameter estimation in <10 seconds. The proposed workflow was reproducible across protocols, volunteers and rescans. DICOM parametric maps were exported from the scanner for further analyses. Comparisons between $NN_{MLE}$ and $NN_{GT}$ suggested $NN_{MLE}$ parameter estimates to be more consistent with conventional fitting, a finding supported by offline evaluations.

**Conclusion:** Real-time, inline parameter estimation with the proposed generalisable framework resolves a key practical barrier to the potential clinical uptake of advanced qMRI methods, enabling their efficient integration into clinical workflows. Next steps include incorporation of pre-processing methods and evaluation in pathology.

# 1 INTRODUCTION

Quantitative MRI (qMRI) enables the estimation of tissue properties of interest using biophysical models that relate these properties to measured MR signals. Compared to conventional MRI, it provides biomarkers that have less dependence on acquisition settings and may better inform on pathology. One such advanced qMRI technique is Neurite Orientation Dispersion and Density Imaging (NODDI), a three-compartment model of multi-shell diffusion MRI signals[1]. As a brain microstructure imaging method, NODDI provides voxel-wise estimates of the orientation dispersion index (ODI, degree of spreading of neurites), the neurite density index (NDI, the intra-neurite fraction of tissue) and the free water fraction (FWF, the volume fraction of bulk liquid undergoing free diffusion), as well as the predominant neurite orientations.

Although promising in research settings[2], qMRI methods have not yet reached "clinical maturity", according to Granziera et al.[3], with multi-shell diffusion models such as NODDI further from adoption than other techniques. One key reason, they believe, is that the methods and tools to reconstruct parametric maps are not clinically available[3]. Indeed, parameter estimation is typically done offline. Often this requires exporting image data from MRI systems onto high-performance research workstations where, depending on model complexity, conventional parameter estimation may take multiple hours[1], though strategies for acceleration exist[4]. Although it has been argued that a lack of accumulated evidence and validation, rather than computational costs, is the primary inhibitor to clinical adoption of techniques like NODDI[2], we believe that the computational costs themselves are a fundamental reason for this lack, restricting large-scale clinical research and validation studies. They also still render many potential applications in acute or point-of-care scenarios categorically infeasible. Even when lengthy reconstruction times are acceptable, the need for external computing resources is disruptive, and parametric maps seldomly re-enter the clinical data stream, failing to reach reporting systems. Recently, 50% of surveyed European neuroradiologists suggested that technical improvements in software and hardware would catalyse greater uptake of qMRI[5]; a lack of processing software (23%) and time intensive processes (39%) were identified as major impediments[5]. Altogether, we believe that the entire conventional qMRI workflow is impractical and incompatible with clinical practice, holding advanced techniques back in 'research mode'. Fortunately, two exciting recent advances with machine learning (ML) promise to make qMRI more clinically viable.

First, ML is revolutionising parameter estimation: in particular, fully-connected neural networks (NN) trained with synthetic data can replace computationally expensive model fitting procedures, producing voxel-wise parameter estimates near instantaneously[6–11]. An additional advantage is the potential to resolve model (parameter) degeneracy[12,13], which occurs when multiple parameter combinations produce indistinguishable noisy signals. ML approaches are typically either supervised[6,9] or self-supervised[10] (unsupervised[8]). The former utilises ground truth (GT) parameters as targets (labels) during training, computing and minimising the loss in parameter space; the latter self-generates labels to minimise the loss in signal space. Supervised learning is a flexible and intuitive approach but may lead to biased parameter estimates, whereas self-supervised learning may yield more accurate parameter estimates, at the cost of implementational challenges[7]. To overcome this trade-off, Epstein et al. recently proposed supervised learning with non-GT target parameters, instead utilising the estimates returned by a conventional fit of the synthetic data[7]. This strategy was shown to emulate the improved accuracy of self-supervised learning, while retaining the scalability and simplicity of a supervised framework.

Second, MRI vendors have recently begun to support inline inference with trained NNs for image manipulation tasks. For example, the Siemens Framework for Image Reconstruction Environments (FIRE) prototype[14] permits a containerised deployment of trained NNs[15,16], interfacing with the reconstruction pipeline using the open-source ISMRM raw data format[17]. Complete integration of trained NNs into the vendors' native image reconstruction pipeline is also possible, using C or C++ runtime libraries[18–21]. Otherwise, external third-party reconstruction frameworks may be configured for inline deployment of trained NNs (e.g. Gadgetron[22] with the InlineAI module[23,24]), however this still demands dedicated computing and research infrastructures, which clinical settings may be unable to accommodate.

The primary aim of this work was to exploit and combine the two outlined advances to achieve real-time, inline, fully scanner-integrated qMRI parameter estimation with ML. We chose to demonstrate this strategy for NODDI, with the secondary aim of applying and validating Epstein et al.'s novel supervised ML qMRI model fitting approach[7] for reduced parameter estimation bias without the complexity of self-supervised learning.

# 2 METHODS

## 2.1 Workflow overview

To introduce the core of our method, Figure 1 illustrates the proposed 'clinical mode' workflow for qMRI, highlighting key differences from the conventional 'research mode' approach, including real-time neural network (NN) based parameter estimation using the scanner's computing hardware, and integration with clinical reporting systems.

Inline NN inference was performed using ONNX (Open Neural Network Exchange) Runtime[25] libraries (https://www.nuget.org/packages/Microsoft.ML.OnnxRuntime), which were compiled into a custom reconstruction program developed in Siemens' Image Calculation Environment (ICE), using C++. The ICE program first accumulates images into a 4D data structure as they pass through the reconstructor as usual, slice-by-slice. Once all images have been reconstructed and the accumulated data structure is complete, volumes are normalised by the mean $b = 0$ s/mm$^2$ signal in each voxel. Inference is then performed voxel-wise, on the reconstruction server's CPU, with a trained NN in the ONNX format, as supplied by the user. The resultant parametric maps of ODI, NDI and FWF are finally output as a separate DICOM series, with floating point numbers stored as 12-bit integers scaled by 1000, achieving a precision of three decimal places. Parametric maps may be visualised, analysed and inspected natively on the console PC and sent directly to reporting systems for clinical evaluation.

The following subsections detail how NNs were trained, deployed in vivo and evaluated.

## 2.2 Neural network (NN) development

NNs were developed and trained offline. Operating on a voxel-wise basis, NNs receive as input a diffusion-encoded signal vector and output a corresponding parameter vector (summarised in Figure 2). This setup enables training using synthetic diffusion MRI signals. First, ground truth (GT) parameter values for orientation dispersion index (ODI), neurite density index (NDI), free water fraction (FWF) were drawn from random uniform distributions[6], $U$, (Figure 2) for $2^{18}$ unique combinations (or training 'datasets'):

$$\mathrm{ODI} \sim U(0.05, 1.0);\ \mathrm{NDI} \sim U(0.05, 1.0);\ \mathrm{FWF} \sim U(0.0, 0.95).$$

Distribution limits were chosen to reduce parameter degeneracies. Each parameter combination was allocated a fibre orientation, drawn from a random uniform distribution over the unit sphere[26]. With each unique combination of GT parameters, a diffusion encoded signal was synthesised with the forward NODDI model using an internal version of the NODDI MATLAB Toolbox (NMT, http://mig.cs.ucl.ac.uk/index.php?n=Tutorial.NODDImatlab). To demonstrate adaptability to multiple protocols, signals were synthesised for two-shell (81 encodings) and three-shell (116 encodings) diffusion schemes (Table S1). Signals were normalised to $b = 0$ s/mm$^2$ and complex Gaussian noise of $\mathrm{SNR} = 15$ at $b = 0$ s/mm$^2$ (representative of white matter [WM] in vivo) was added before calculating the magnitude, resulting in Rician-distributed measurements.

Training data were fitted conventionally with a maximum likelihood estimator (MLE) using the NMT. To accelerate fitting, the grid search across parameter space was skipped, directly supplying GTs to initialise the gradient descent algorithm. Following Epstein et al.[7], MLE parameters were then utilised as training labels. For comparison, a separate supervised NN was also trained with GT training labels. The two NNs are herein referred to as $NN_{MLE}$ and $NN_{GT}$. They were implemented in PyTorch[27] (version 2.2.1) with the fully connected architecture illustrated in Figure 2 and summarised in Table S2, together with training parameters. The output layer utilised a hard sigmoid activation function to constrain the NN output between 0 and 1. After training, the best model, with lowest validation loss, was saved for evaluation.

## 2.3 In vivo demonstration

Full online operation was tested on a MAGNETOM Vida 3T system (Siemens Healthineers, Forchheim, Germany). With informed consent and approval of the local ethics committee, two healthy volunteers (V1: male, 31y; V2: male, 25y) were imaged using both diffusion protocols in Table S1. V2 was rescanned without repositioning, yielding a total of 6 diffusion imaging datasets. After transferring the ONNX files of trained NNs onto the console PC, the custom ICE program was used to reconstruct the raw data with trained NNs inline. Parametric maps and diffusion encoded images were exported in DICOM format. After offline conversion to NIFTI format, parametric maps were also reconstructed on a research workstation with a conventional MLE routine using the NMT, without any additional pre-processing. For tissue specific quantitative analyses, masks for each image series were obtained utilising SynthSeg[28] on mean $b = 0$ s/mm$^2$ images and thresholding the probabilistic mask at 0.95 for WM, grey matter (GM) and CSF (ventricles).

### 2.4 Synthetic NN evaluation

To address the secondary aim of this work, trained NNs were evaluated with synthetic test data. 3D parameter space was sampled uniformly within 0.05–0.95 for ODI, NDI and FWF. GT parameters were spaced at increments of 0.1, giving $P = 1000$ unique parameter combinations, as shown in Figure 2. Then, to investigate estimation accuracy and variability, test signals were synthesised with $N = 100$ unique noise realisations for each parameter combination, for a total of 100000 test 'datasets'. To control for variations due to the fibre orientation, it was fixed to $[x, y, z] = [1,0,0]$.

To prevent degenerate parameter estimates (e.g. ODI and NDI for high FWF) from skewing or biasing quantitative analyses, a degeneracy-free subspace, representative of tissue (WM and GM), was defined: 0.05–0.55 (ODI), 0.35–0.85 (NDI), 0.05–0.45 (FWF). Model performance was evaluated at each GT parameter grid point, $x_p = [\mathrm{ODI}, \mathrm{NDI}, \mathrm{FWF}]$, by computing the mean, $\mu_p$, bias, $\boldsymbol{b}_p$, and standard deviation, $\sigma_p$, of fitted parameters across repeats:

$$\mu_p = \frac{1}{N}\sum_{n=1}^{N} \hat{x}_{pn}\,; \boldsymbol{b}_p = \mu_p - x_p; \sigma_p = \sqrt{\frac{\sum_{n=1}^{N}\left(\hat{x}_{pn} - \mu_p\right)^2}{N}},$$

where $\hat{x}_{np}$ are the estimated parameters of noise realisation $n$ and grid point $p$. As summary metrics, the mean of magnitude biases, $|\boldsymbol{b}|$, and the mean of the absolute differences ($\overline{|\Delta|}$) of paired NN-MLE estimates from a conventional fit, were calculated as follows:

$$\overline{|\boldsymbol{b}|} = \frac{1}{P}\sum_{p=1}^{P}|\boldsymbol{b}_p|;\ \overline{|\Delta|} = \frac{1}{PN}\sum_{p=1}^{P}\sum_{n=1}^{N}\left|\hat{x}_{np}^{\mathrm{NN}} - \hat{x}_{np}^{\mathrm{MLE}}\right|.$$

# 3 RESULTS

In vivo NODDI parametric maps were estimated inline and displayed on the scanner console PC, demonstrating successful integration of the NN into the scanner's reconstruction environment (Figure 3). The inline reconstruction for the entire 4D dataset took under 10 seconds.

In vivo parametric maps were exported in DICOM format for analysis. They are displayed in Figure 4 for V2 and the two-shell protocol (see Figure S1 for the three-shell protocol) and compared to parametric maps obtained with a conventional MLE fit of an axial slice. On a research workstation equipped with an Intel Xeon w3-2435 CPU, the MLE fit took approximately 3 minutes for an individual masked slice, using 6 parallel processes. This scales to approximately 2 hours for a whole, masked brain (~0.04s per voxel). Parametric maps for all volunteers, rescans and protocols, as estimated with $NN_{MLE}$, are shown in Figure S2.

Also displayed in Figure 4 are scatter plots of estimated parameters in different tissue types (combining all volunteers and rescans, two-shell protocol), comparing conventional parameter estimation to $NN_{MLE}$ and $NN_{GT}$. A complete reference of tissue-wise parameter estimates for all volunteers and rescans, protocols and estimation methods is provided in Table S3. $NN_{MLE}$ returns slightly less biased estimates, more consistent with MLE, in WM (NDI, FWF), GM (ODI, FWF) and CSF (FWF), as substantiated by smaller mean absolute differences between paired NN-MLE estimates for $NN_{MLE}$. Moreover, Figure 4 is reproduced for the three-shell protocol in Figure S1, with similar but slightly improved patterns of bias and agreement of $NN_{MLE}$ and $NN_{GT}$ (in WM).

Figure 5 summarises the distributions of tissue-wise parameter estimates across volunteers and rescans. Estimates appear in general agreement, and small deviations are consistent across methods (e.g. slightly increased NDI in WM for V2 rescan). ODI, FWF and WM appear more stable than NDI and GM. Figure 5 is reproduced for the three-shell protocol in Figure S3.

The offline evaluation of trained NNs with synthetic test data is summarised in Figure S4, showing vector plots of the bias and line plots of the standard deviation of estimated test parameters, as well as scatter plots comparing MLE to $NN_{MLE}$ and $NN_{GT}$ estimates, all within a tissue parameter subspace. $NN_{MLE}$ and MLE bias vectors closely overlap, indicating comparable estimation biases, whereas $NN_{GT}$ bias vectors extend and deviate more from the GT. Mean magnitude biases ($\overline{|\boldsymbol{b}|}$) were 0.029, 0.029 and 0.068 (Figure S4) for MLE, $NN_{MLE}$ and $NN_{GT}$ respectively. In scatter plots, $NN_{GT}$ also returns a greater point spread, away from the line of identity, as reflected quantitatively by increased mean absolute differences. The same visualisations are reproduced for the three-shell protocol in Figure S5. Additional diffusion encodings reduce overall bias, as well as the difference between $NN_{MLE}$ and $NN_{GT}$.

# 4 DISCUSSION

This study successfully demonstrated the possibility of scanner-integrated, real-time qMRI by leveraging ML methods, resolving a key practical barrier inhibiting the uptake of advanced qMRI in clinics. With the proposed framework, parametric maps can be incorporated into clinical workflows and reporting systems, so that advanced qMRI could form a part of multimodal medical diagnoses and monitoring, without requiring cumbersome offline processing and data exports or imports. Although other existing methodologies could support this (e.g. Gadgetron[22,23]), they are often still considered research tools, requiring additional computing infrastructures. With a repository of pretrained NNs, our strategy has no additional demand for local hardware resources or specialised expertise, offering the closest possible and least disruptive clinical integration of advanced qMRI. Further, immediate delivery of quantitative parametric maps enables new applications for such techniques in acute and point-of-care settings. While NNs are independent of generic acquisitions parameters (e.g. spatial resolution or TR, although SNR should be similar between training and in vivo data), they depend on the diffusion encoding scheme, or generally, model-related protocol parameters. Therefore, a new diffusion MRI protocol requires training a new NN, though the ICE program is generalised to any diffusion MRI protocol and does not require recompilation. Providing a new NN would be akin to uploading new files of b-vectors and b-values onto the scanner PC. As NN training is relatively fast and utilises synthetic data, it is not a burdensome procedure; indeed, scanner-integrated training of $NN_{GT}$ networks could be a future way of streamlining new protocol generation. However, the ICE program does not immediately generalise to any arbitrary qMRI modality; this would require source code modifications specific to the desired application.

Visual inspection of in vivo parametric maps shows similar outputs for NN and conventional MLE estimates, although minor quantitative differences, e.g. increased NN FWF in white matter, are detectable. While overestimation of FWF is evident in scatter plots of Figure 4, it is less apparent in NN evaluations with synthetic data in Figure S4, especially for $NN_{MLE}$, suggesting that either mismatches between training and in vivo SNR, or 'out-of-model' effects, like data quality issues, cause this discrepancy. The real-world manifestation of noise is only partially addressed by training on Rician-distributed measurements, since multi-channel array coils and parallel imaging typically result in more complex noise behaviours, such as spatially varying noise levels[29]. While utilising the adaptive combine algorithm for channel combination results in Rician-distributed measurements[30], the 'SENSE1' combination strategy[31] (not available to us) may prove to be more optimal for our application. The discrepancy of FWF also improves slightly for the three-shell protocol with additional diffusion encodings.

Otherwise, in vivo overestimations of ODI, or of NDI by $NN_{GT}$, also manifest in scatter plots of synthetic data estimates in Figure S4, suggesting that they are of systematic origin. Further refinement of training data distributions, NN architecture and settings must take place to reduce these biases. As training the NN involves balancing accuracy across three interdependent parameters, and gains in one may lead to losses in another, one strategy might involve training separate networks to estimate each parameter. Crucially, in vivo MLE estimates serve only as benchmark or reference, not as ground truth. For discrepancies that do not manifest in synthetic data, it cannot be concluded whether MLE or NN estimates are more accurate.

Diffusion MRI has been known for its sensitivity to image artifacts and data quality issues, like geometric distortions and misalignments, caused by subject motion, magnetic susceptibility and eddy current effects[32]. Eddy currents are considered more problematic for diffusion modelling, as they directly arise from the application of strong diffusion gradients, and therefore produce variable misalignments across diffusion encodings[33]. In our case, this could lead to 'out-of-model' effects unseen to NNs. However, on modern scanner systems with state-of-the-art eddy current compensation strategies and the highest possible EPI readout bandwidths, even $b = 2000$ s/mm$^2$ images appear minimally corrupted by eddy currents. Shorter scan times achieved with acceleration techniques have reduced motion artifacts, too. Still, protocols for highly advanced diffusion models, such as soma and neurite density imaging (SANDI)[34], may require longer acquisitions and stronger diffusion encodings, at which stage residual eddy current effects and motion must be corrected for in post-processing. With this proof-of-concept study, we demonstrate the capabilities of a minimally-pre-processed workflow which may be adequate for certain applications. To cover more use cases in the future, we plan to integrate as many pre-processing steps as possible into a state-of-the-art inline processing pipeline. A number of groups have already proposed susceptibility distortion[35,36], head motion and eddy-current distortion correction techniques[37] based on deep learning, which promise to be fast enough for inline implementation. It is already possible to perform inline denoising using vendor-provided methods, such as Deep Resolve Gain.

From evaluations of synthetic test data in Figures S4 and S5, it is clear that $NN_{MLE}$ produces estimates closer to the GT and to MLE estimates than $NN_{GT}$. This is in agreement with Epstein et al., who proposed and demonstrated the MLE-label framework for the intravoxel incoherent motion (IVIM) model[7]. We report that the strategy extends to more complex microstructure models, like NODDI, although estimation bias must be interpreted cautiously in the presence of parameter degeneracies. The framework offers a powerful and practical ML solution that emulates conventional fitting, with smaller additional biases. This is particularly crucial for atypical parameter distributions in disease.

# 5 CONCLUSION

In a proof-of-principle work, we demonstrate that real-time, inline NODDI is possible with scanner-integrated ML. Further, we have shown that a novel strategy to reduce parameter estimation bias and emulate conventional fitting with supervised ML extends well to a complex model like NODDI. These powerful strategies should generalise to any qMRI model and scanner vendor offering development capabilities on their reconstruction platform. This work represents a significant step in bringing qMRI techniques closer to clinical utility, delivering full integration of parametric maps with clinical workflows. The next steps will involve inline incorporation of common pre-processing steps, and demonstration and validation in pathology. Outside of the clinic, real-time data quality monitoring can be impactful even in research settings, accelerating and streamlining protocol development and optimisation.

## DATA AVAILABILITY STATEMENT

Code to support and reproduce this work relies on an internal version of the NODDI MATLAB Toolbox (NMT) and will be made publicly available as part of the next NMT release. Additional methodological details are available from the authors upon reasonable request.

## ACKNOWLEDGEMENTS

This work received support from the following funding sources:

- EPSRC Impact Acceleration Account to UCL 2022-26 (EP/X525649/1) [SR, DT, DA, HZ].
- NIH (1R01MH130362) [SR, HZ].
- UCLH NIHR Biomedical Research Centre [JST, SR] and the EPSRC-funded UCL Centre for Doctoral Training in Intelligent, Integrated Imaging in Healthcare (i4health) (EP/S021930/1) [SR].
- UKRI Future Leaders Fellowship (MR/S032290/1) [JS].

## FIGURES

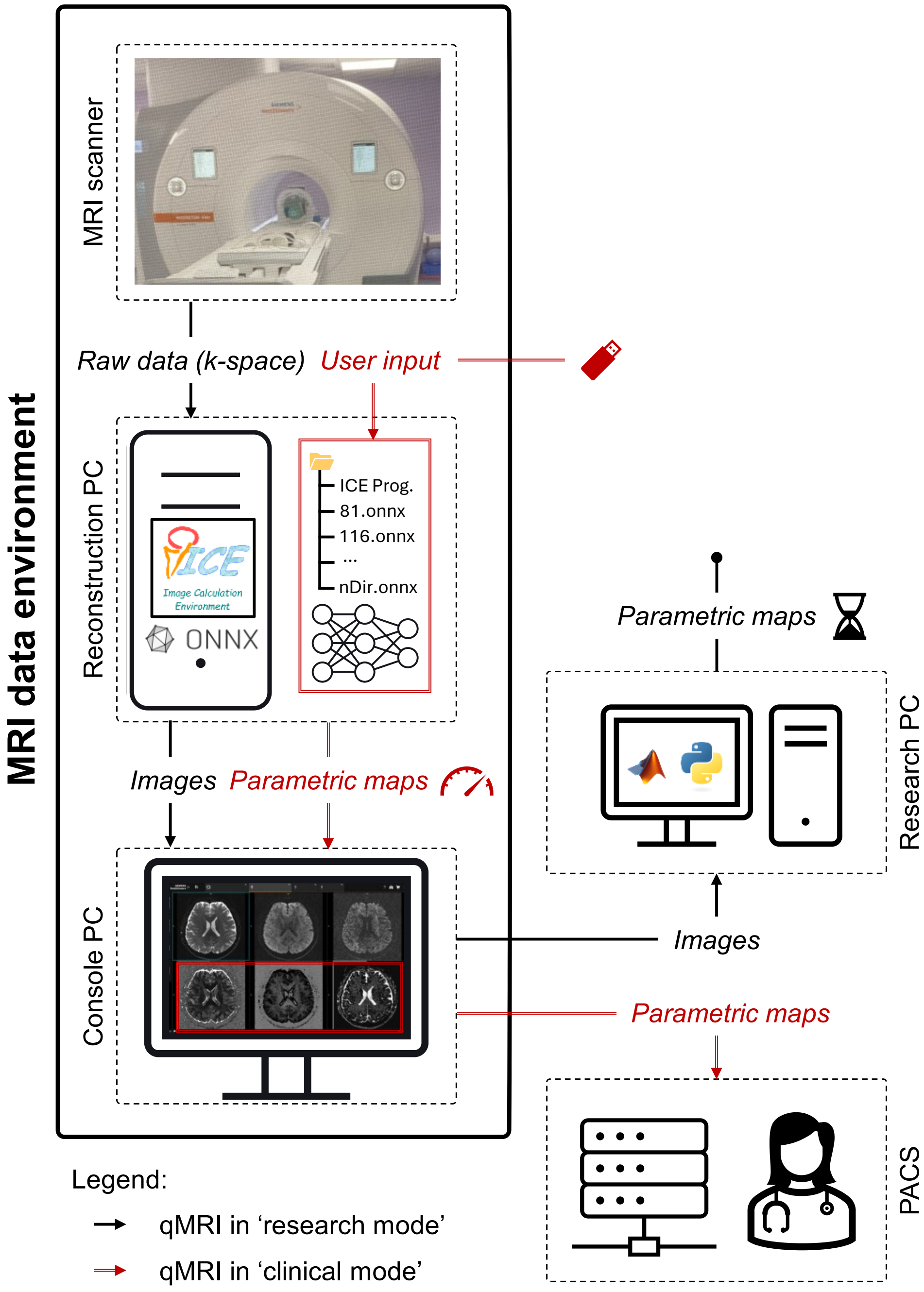

### Figure 1

A summary schematic of the proposed workflow for qMRI in 'clinical mode', contrasted against a conventional workflow for qMRI in 'research mode'. Raw k-space data are sent from the scanner to the vendor's reconstruction server and reconstructed into images slice by slice. From this point, the two workflows differ. In 'research mode' qMRI: images are displayed on the scanner console PC and exported for offline qMRI parameter estimation, usually onto a high-performance research workstation, to meet the high computational demands of parameter fitting. Once estimated, parametric maps often remain within the research environment, as a transfer back onto the scanner console PC is cumbersome. In the proposed ‘clinical mode’ qMRI, still on the reconstruction server, slices and multiple image data dimensions are accumulated and fed through an integrated trained neural network (NN) for real-time inference of qMRI model parameters. The resulting parametric maps are stored in DICOM format and sent to the console PC for display. From there, they may be forwarded to PACS, or other clinical reporting systems, alongside conventional diagnostic imaging data. A one-time user import of ICE program files and ONNX trained NNs is required, e.g. via flash drive. A simple NN naming scheme, like using the number of diffusion encodings, allows automatic selection and concurrent storage of different NNs.

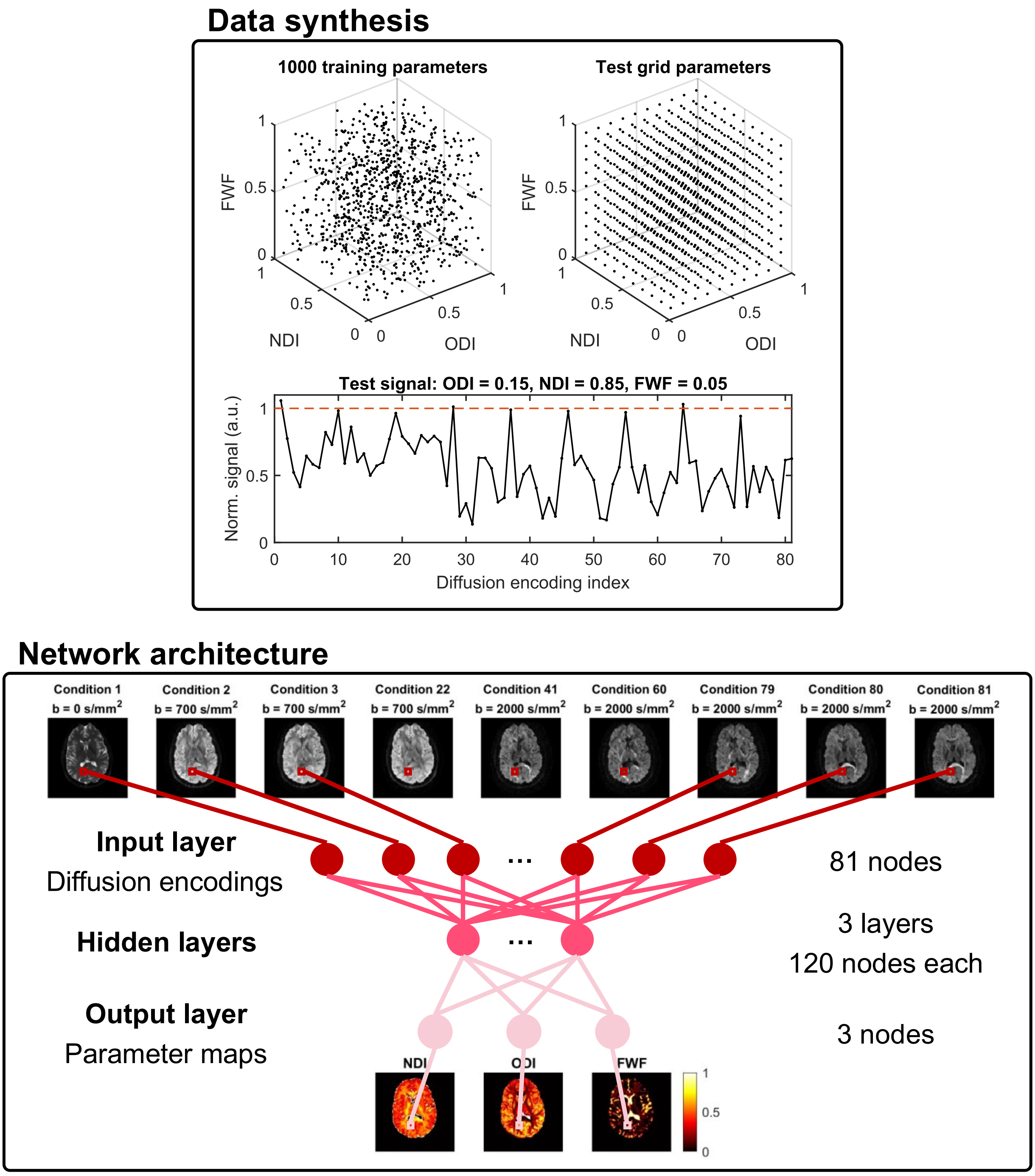

## Figure 2

The upper panel shows training and test data parameters in 3D parameter space of orientation dispersion index (ODI), neurite density index (NDI), free water fraction (FWF). For each grid point of test parameters, 100 signals with unique noise were synthesised, with an example shown for typical white matter parameters (high NDI, low ODI and FWF). The lower panel shows a schematic of the fully connected neural network (NN) architecture, employed in a voxel-wise manner, and demonstrated for the two-shell protocol (the number of input notes is protocol dependent). Images in the schematic are from example data of the NODDI MATLAB Toolbox.

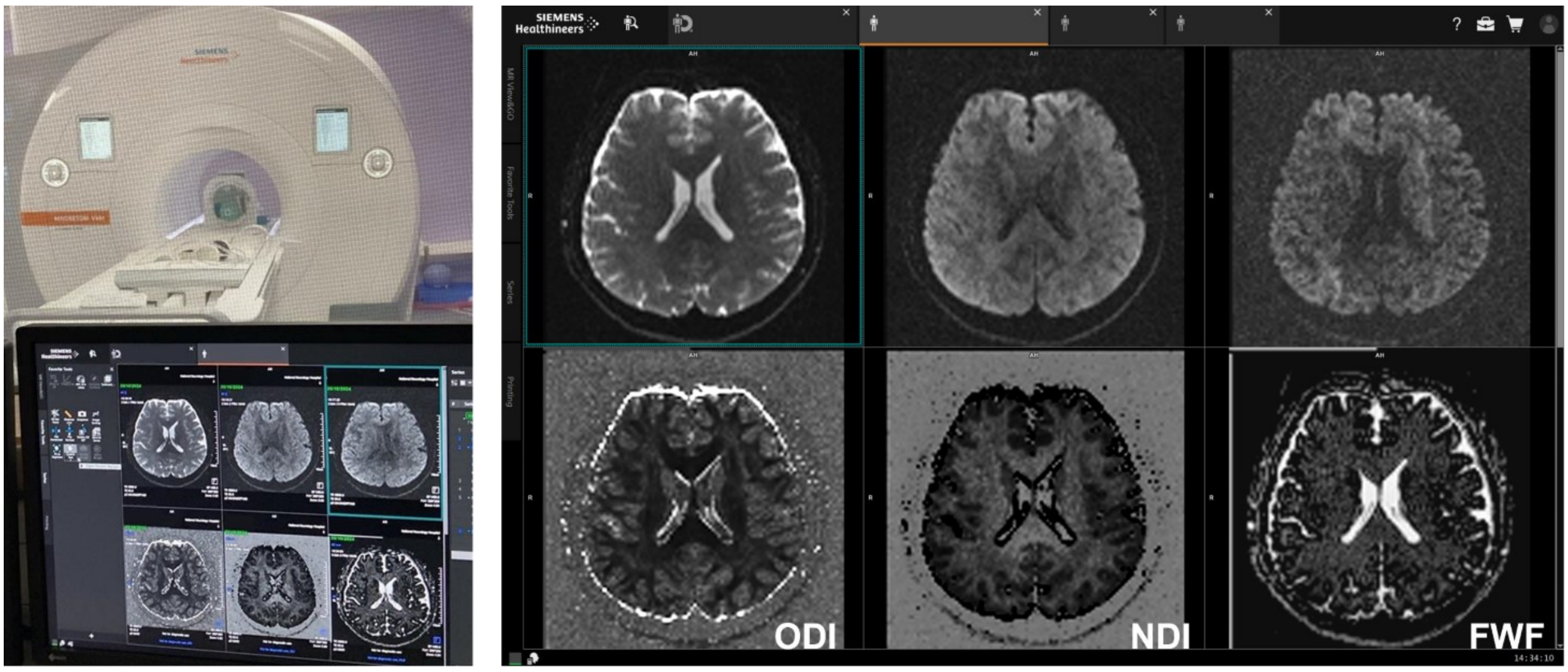

## Figure 3

Right, a screenshot of the MAGNETOM Vida 3T (Siemens Healthineers, Forchheim, Germany) console monitor, showing inline reconstructed maps (ODI, NDI, FWF from left to right, bottom row) for a participant scanned. Sample diffusion encoded images are shown in the top row. Left shows a photograph of the console from the control room.

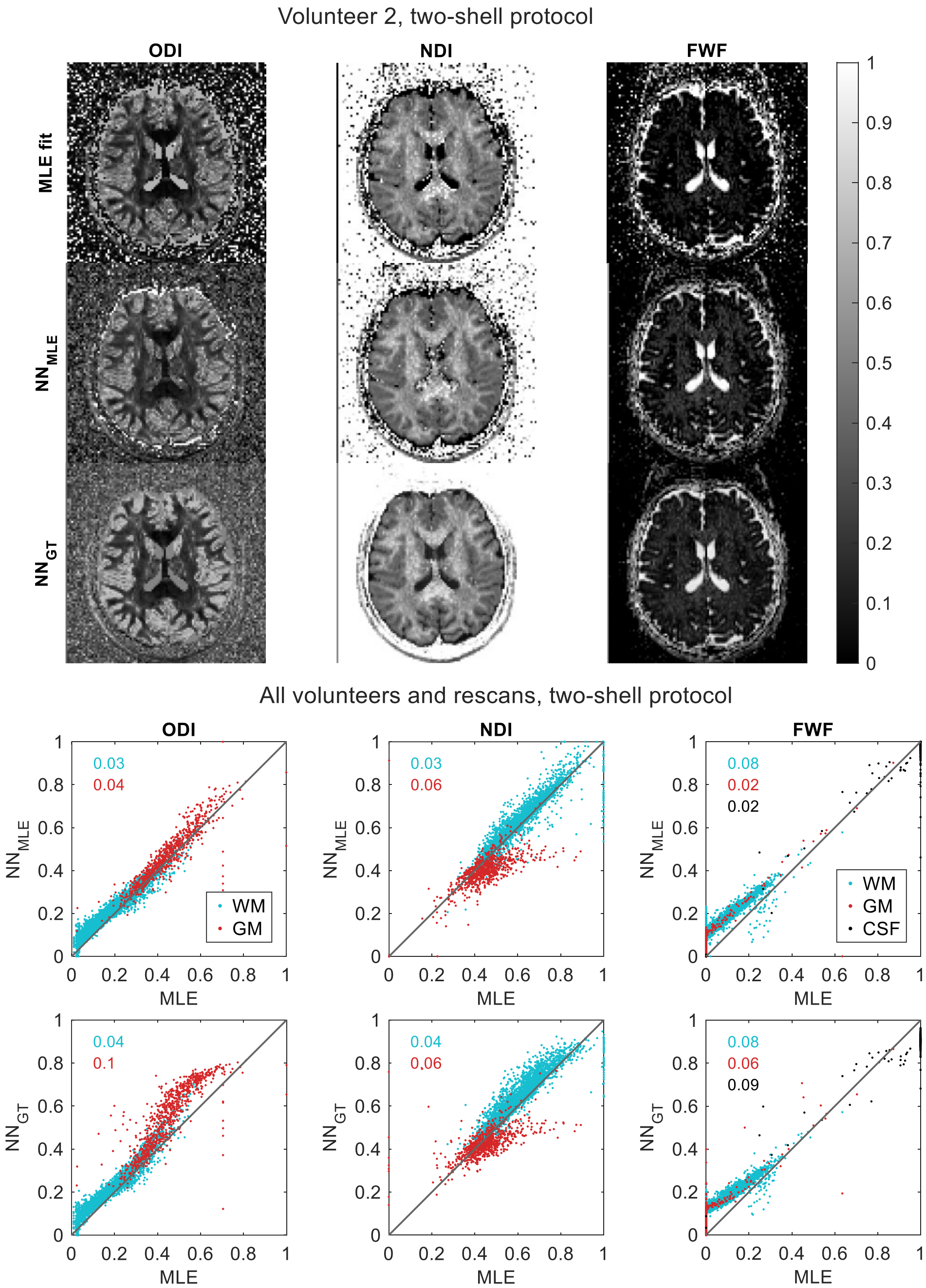

## Figure 4

Estimated orientation dispersion index (ODI), neurite density index (NDI) and free water fraction (FWF) parameter maps for an axial slice of V2 and the two-shell protocol. The upper row shows maps fitted conventionally with the NODDI MATLAB Toolbox (indicated MLE); the next two rows show maps inferred with the two trained neural networks ($NN_{MLE}$ and $NN_{GT}$). The bottom two rows show scatter plots of estimated parameters for each tissue type of single axial slices of all volunteers and rescans using the two-shell protocol, comparing MLE to $NN_{MLE}$ (row 4) and $NN_{GT}$ (row 5). Pairwise MLE-NN mean absolute differences for each tissue class are noted in the top left of each panel. NN parameter maps were exported as DICOMs, converted to NIFTI format and rescaled from integer (range 0 to 1000) to float (range 0 to 1). No post-processing was performed.

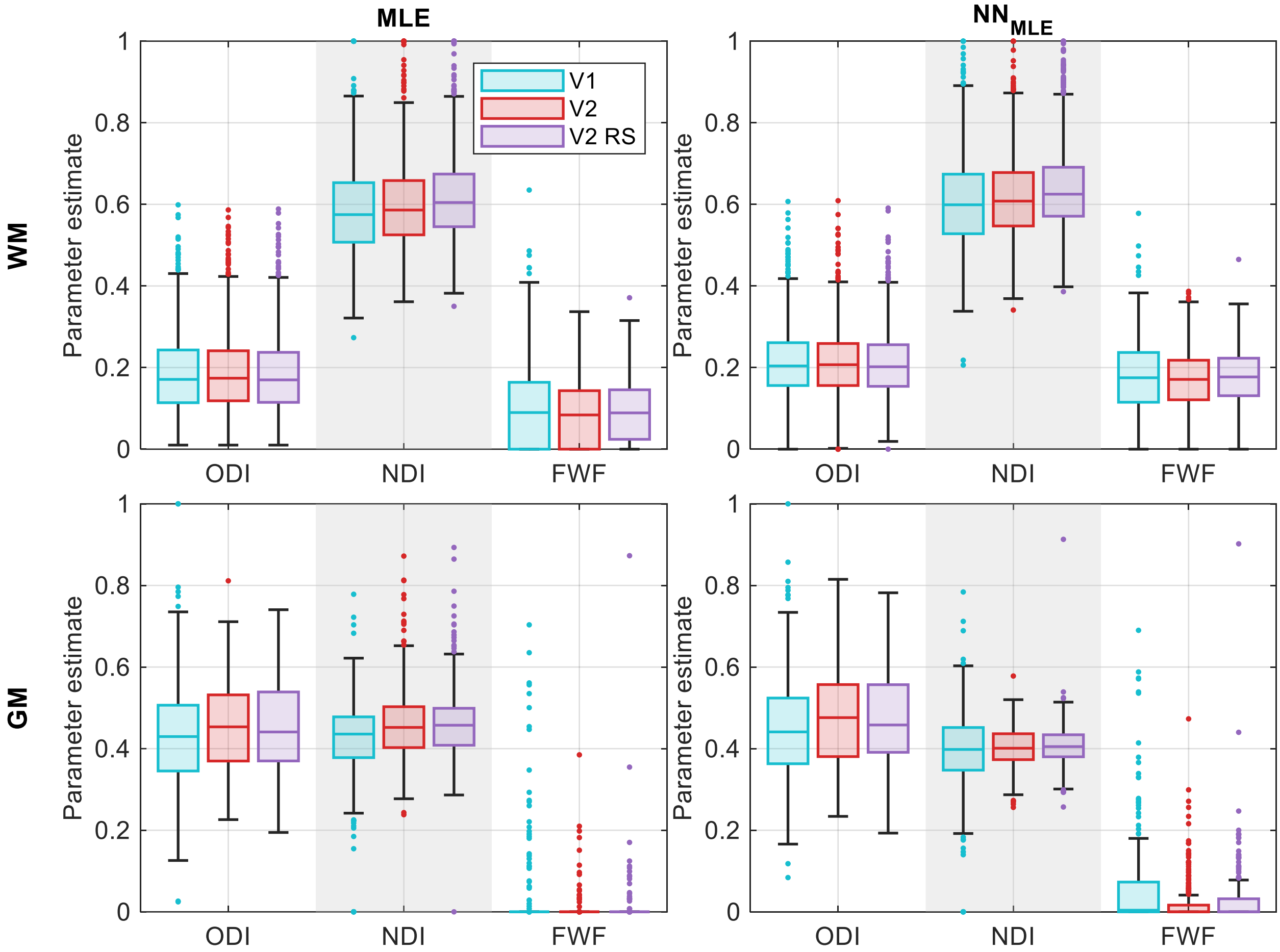

## Figure 5

Boxplots of estimated orientation dispersion index (ODI), neurite density index (NDI) and free water fraction (FWF) parameters for an axial slice across all volunteers and rescans (different colours) of the two-shell protocol. Each panel shows data for a particular tissue type and estimation method, as indicated in row and column labels. The solid horizontal line indicates the median, the box indicates the interquartile range (IQR) and whiskers indicate outlier bounds, 1.5 IQRs from the lower and upper quartiles.

# REFERENCES

1. Zhang H, Schneider T, Wheeler-Kingshott CA, Alexander DC. NODDI: Practical in vivo neurite orientation dispersion and density imaging of the human brain. *NeuroImage*. 2012;61(4):1000-1016. doi:10.1016/j.neuroimage.2012.03.072

2. Kamiya K, Hori M, Aoki S. NODDI in clinical research. *Journal of Neuroscience Methods*. 2020;346:108908-108908. doi:10.1016/j.jneumeth.2020.108908

3. Granziera C, Wuerfel J, Barkhof F, et al. Quantitative magnetic resonance imaging towards clinical application in multiple sclerosis. *Brain*. 2021;144(5):1296-1311. doi:10.1093/brain/awab029

4. Daducci A, Canales-Rodríguez EJ, Zhang H, Dyrby TB, Alexander DC, Thiran JP. Accelerated Microstructure Imaging via Convex Optimization (AMICO) from diffusion MRI data. *NeuroImage*. 2015;105:32-44. doi:10.1016/j.neuroimage.2014.10.026

5. Manfrini E, Smits M, Thust S, et al. From research to clinical practice: a European neuroradiological survey on quantitative advanced MRI implementation. *European Radiology*. 2021;31(8):6334-6341. doi:10.1007/s00330-020-07582-2

6. Gyori NG, Palombo M, Clark CA, Zhang H, Alexander DC. Training data distribution significantly impacts the estimation of tissue microstructure with machine learning. *Magnetic Resonance in Medicine*. 2022;87(2):932-947. doi:10.1002/mrm.29014

7. Epstein SC, Bray TJP, Hall-Craggs M, Zhang H. Choice of training label matters: how to best use deep learning for quantitative MRI parameter estimation. *Machine Learning for Biomedical Imaging*. 2024;2(January 2024):586-610. doi:10.59275/j.melba.2024-geb5

8. Barbieri S, Gurney-Champion OJ, Klaassen R, Thoeny HC. Deep learning how to fit an intravoxel incoherent motion model to diffusion-weighted MRI. *Magnetic Resonance in Medicine*. 2020;83(1):312-321. doi:10.1002/mrm.27910

9. de Almeida Martins JP, Nilsson M, Lampinen B, et al. Neural networks for parameter estimation in microstructural MRI: Application to a diffusion-relaxation model of white matter. *NeuroImage*. 2021;244:118601-118601. doi:10.1016/j.neuroimage.2021.118601

10. Sen S, Singh S, Pye H, et al. ssVERDICT: Self-supervised VERDICT-MRI for enhanced prostate tumor characterization. *Magnetic Resonance in Medicine*. 2024;92(5):2181-2192. doi:10.1002/mrm.30186

11. Bray TJ, Minore GV, Bainbridge A, et al. RAIDER: Rapid, anatomy-independent, deep learning-based PDFF and R2* estimation using magnitude-only signals, dual neural networks and training data distribution design. *Machine Learning for Biomedical Imaging*. 2025;3(October 2025):521-544. doi:10.59275/j.melba.2025-bac4

12. Guerreri M, Epstein S, Azadbakht H, Zhang H. Resolving Quantitative MRI Model Degeneracy with Machine Learning via Training Data Distribution Design. In: Frangi A, De Bruijne M, Wassermann D, Navab N, eds. *Information Processing in Medical Imaging*. Vol 13939. Lecture Notes in Computer Science. Springer Nature Switzerland; 2023:3-14. doi:10.1007/978-3-031-34048-2_1

13. Minore GV, Dwyer-Hemmings L, Bray TJP, Zhang H. Resolving Quantitative MRI Model Degeneracy in Self-supervised Machine Learning. In: Oguz I, Zhang S, Metaxas DN, eds.

*Information Processing in Medical Imaging*. Vol 15830. Lecture Notes in Computer Science. Springer Nature Switzerland; 2026:186-199. doi:10.1007/978-3-031-96625-5_13

14. Chow K, Kellman P, Xue H. Prototyping image reconstruction and analysis with FIRE. In: *SCMR 24th Annual Scientific Sessions*. 2021.

15. Yoon S, Nakamori S, Amyar A, et al. Accelerated Cardiac MRI Cine with Use of Resolution Enhancement Generative Adversarial Inline Neural Network. *Radiology*. 2023;307(5). doi:10.1148/radiol.222878

16. Vornehm M, Wetzl J, Giese D, et al. CineVN: Variational network reconstruction for rapid functional cardiac cine MRI. *Magnetic Resonance in Medicine*. 2025;93(1):138-150. doi:10.1002/mrm.30260

17. Inati SJ, Naegele JD, Zwart NR, et al. ISMRM Raw data format: A proposed standard for MRI raw datasets. *Magnetic Resonance in Medicine*. 2017;77(1):411-421. doi:10.1002/mrm.26089

18. Yun SM, Hong SB, Lee NK, et al. Deep learning-based image reconstruction for the multi-arterial phase images: improvement of the image quality to assess the small hypervascular hepatic tumor on gadoxetic acid-enhanced liver MRI. *Abdominal Radiology*. 2024;49(6):1861-1869. doi:10.1007/s00261-024-04236-5

19. Jung HK, Choi Y, Kim S, Nickel D, Park JE, Kim HS. Image quality assessment and white matter hyperintensity quantification in two accelerated high-resolution 3D FLAIR techniques: Wave-CAIPI and deep learning–based SPACE. *Clinical Radiology*. 2025;82:106783-106783. doi:10.1016/j.crad.2024.106783

20. Wei H, Yoon JH, Jeon SK, et al. Enhancing gadoxetic acid–enhanced liver MRI: a synergistic approach with deep learning CAIPIRINHA-VIBE and optimized fat suppression techniques. *European Radiology*. 2024;34(10):6712-6725. doi:10.1007/s00330-024-10693-9

21. Rot S, Dragonu I, Thomas D, Alexander DC, Zhang H. Real-time quantitative MRI enabled by scanner integrated machine learning: a proof of principle with NODDI. In: *Proceedings of the International Society of Magnetic Resonance in Medicine*. 2025:0340.

22. Hansen MS, Sørensen TS. Gadgetron: An open source framework for medical image reconstruction. *Magnetic Resonance in Medicine*. 2013;69(6):1768-1776. doi:10.1002/mrm.24389

23. Xue H, Davies R, Hansen D, et al. Gadgetron Inline AI: Effective Model inference on MR scanner. In: *Proceedings of the International Society of Magnetic Resonance in Medicine, 27*. 2019:4837.

24. Xue H, Artico J, Fontana M, Moon JC, Davies RH, Kellman P. Landmark Detection in Cardiac MRI by Using a Convolutional Neural Network. *Radiology: Artificial Intelligence*. 2021;3(5):e200197-e200197. doi:10.1148/ryai.2021200197

25. ONNX Runtime developers. ONNX Runtime. Published online 2021. https://onnxruntime.ai/

26. Muller ME. A note on a method for generating points uniformly on n -dimensional spheres. *Communications of the ACM*. 1959;2(4):19-20. doi:10.1145/377939.377946

27. Paszke A, Gross S, Massa F, et al. PyTorch: an imperative style, high-performance deep learning library. In: *Proceedings of the 33rd International Conference on Neural Information Processing Systems*. Curran Associates Inc.; 2019.

28. Billot B, Greve DN, Puonti O, et al. SynthSeg: Segmentation of brain MRI scans of any contrast and resolution without retraining. *Medical Image Analysis*. 2023;86:102789-102789. doi:10.1016/j.media.2023.102789

29. Aja-Fernández S, Vegas-Sánchez-Ferrero G, Tristán-Vega A. Noise estimation in parallel MRI: GRAPPA and SENSE. *Magnetic Resonance Imaging*. 2014;32(3):281-290. doi:10.1016/j.mri.2013.12.001

30. Sakaie K, Lowe M. Retrospective correction of bias in diffusion tensor imaging arising from coil combination mode. *Magnetic Resonance Imaging*. 2017;37:203-208. doi:10.1016/j.mri.2016.12.004

31. Sotiropoulos SN, Moeller S, Jbabdi S, et al. Effects of image reconstruction on fiber orientation mapping from multichannel diffusion MRI: Reducing the noise floor using SENSE. *Magnetic Resonance in Medicine*. 2013;70(6):1682-1689. doi:10.1002/mrm.24623

32. Le Bihan D, Poupon C, Amadon A, Lethimonnier F. Artifacts and pitfalls in diffusion MRI. *Journal of Magnetic Resonance Imaging*. 2006;24(3):478-488. doi:10.1002/jmri.20683

33. Pierpaoli C. Artifacts in Diffusion MRI. In: *Diffusion MRI*. Oxford University Press; 2010:303-318. doi:10.1093/med/9780195369779.003.0018

34. Palombo M, Ianus A, Guerreri M, et al. SANDI: A compartment-based model for non-invasive apparent soma and neurite imaging by diffusion MRI. *NeuroImage*. 2020;215:116835-116835. doi:10.1016/j.neuroimage.2020.116835

35. Duong STM, Phung SL, Bouzerdoum A, Schira MM. An unsupervised deep learning technique for susceptibility artifact correction in reversed phase-encoding EPI images. *Magnetic Resonance Imaging*. 2020;71:1-10. doi:10.1016/j.mri.2020.04.004

36. Legouhy A, Graham M, Guerreri M, et al. Correction of Susceptibility Distortion in EPI: A Semi-supervised Approach with Deep Learning. In: Cetin-Karayumak S, Christiaens D, Figini M, et al., eds. *Computational Diffusion MRI*. Vol 13722. Lecture Notes in Computer Science. Springer Nature Switzerland; 2022:38-49. doi:10.1007/978-3-031-21206-2_4

37. Legouhy A, Callaghan R, Stee W, Peigneux P, Azadbakht H, Zhang H. Eddeep: Fast Eddy-Current Distortion Correction for Diffusion MRI with Deep Learning. In: Linguraru MG, Dou Q, Feragen A, et al., eds. *Medical Image Computing and Computer Assisted Intervention – MICCAI 2024*. Vol 15002. Lecture Notes in Computer Science. Springer Nature Switzerland; 2024:152-161. doi:10.1007/978-3-031-72069-7_15

# SUPPORTING INFORMATION

## Table S1

A summary of the diffusion MRI protocol parameters utilised for data synthesis and the in vivo imaging experiments. The first protocol utilises the example two-shell diffusion scheme of the NODDI MATLAB Toolbox. Diffusion vectors were distributed isotropically, $b = 0$ s/mm$^2$ conditions were interleaved throughout and shells were sampled separately and consecutively in time. The second protocol is a three-shell diffusion scheme with isotropically distributed diffusion vectors and interleaving of diffusion shells in time.

| **Multi-shell diffusion imaging protocols** | |
|---|---|
| **Sequence** | EPI spin echo |
| ***b* values (nr. encodings) (s/mm$^2$)** | |
| **Two-shell protocol** | 0 (9), 700 (24), 2000 (48) |
| **Three-shell protocol** | 0 (12), 300 (8), 700 (32), 2000 (64) |
| **TE (ms)** | 83 |
| **TR (ms)** | 2900 |
| **FOV (mm$^2$)** | 200x200 |
| **Voxel size (mm$^2$)** | 2x2 |
| **Nr. slices** | |
| **Two-shell protocol** | 40 |
| **Three-shell protocol** | 50 |
| **Slice thickness (mm)** | 2 |
| **Bandwidth (Hz/px)** | 2000 |
| **GRAPPA factor, reference lines** | 2, 20 |
| **Simultaneous multi-slice (SMS) factor** | 2 |
| **Partial Fourier** | 7/8 |
| **Gradient scheme** | Monopolar |
| **Scan time (mm:ss)** | |
| **Two-shell protocol** | 04:09 |
| **Three-shell protocol** | 05:52 |
| **Channel combination** | Adaptive combine |

## Table S2

A summary of neural network (NN) architecture and training settings.

| NN architecture | |
|---|---|
| **Network type** | Fully connected feed-forward NN |
| **Input nodes** | 81 or 116 |
| **Hidden layers** | 3 |
| **Hidden nodes** | 120 (per layer) |
| **Hidden activation** | ELU (α = 0.05) |
| **Output nodes** | 3 |
| **Output activation** | Hard sigmoid |
| **Trainable parameters** | 39243 or 43443 |
| **NN training** | |
| **Python library** | PyTorch (version 2.1.1) |
| **Batch size** | 1000 |
| **Epochs** | 500 |
| **Validation fraction** | 0.1 |
| **Loss function** | Mean squared error |
| **Optimiser** | Adam |
| **Learning rate** | 5e-4 |
| **L2 regularisation strength** | 1e-6 |
| **Momentum** | 0.9 |
| **GPU chip** | NVIDIA GeForce RTX 4090 (CUDA 12.1) |
| **Training time (mm:ss)** | ~ 10:00 |

## Table S3

Brain-regional means ($\mu$) and standard deviations ($\sigma$) of in vivo estimated orientation dispersion index (ODI), neurite density index (NDI) and free water fraction (FWF) for all estimation methods (conventional MLE, $NN_{MLE}$, $NN_{GT}$), volunteers, rescans (RS) and protocols.

| Subject | Protocol (nr. shells) | Method | Tissue | Nr. voxels | ODI $\mu \pm \sigma$ | NDI $\mu \pm \sigma$ | FWF $\mu \pm \sigma$ |
|---|---|---|---|---|---|---|---|
| V1 | 2 | MLE | WM | 1022 | 0.18 ± 0.10 | 0.59 ± 0.11 | 0.10 ± 0.09 |
| V2 | 2 | MLE | WM | 1289 | 0.18 ± 0.10 | 0.60 ± 0.11 | 0.09 ± 0.08 |
| V2 RS | 2 | MLE | WM | 1285 | 0.18 ± 0.10 | 0.62 ± 0.11 | 0.09 ± 0.08 |
| V1 | 3 | MLE | WM | 1022 | 0.18 ± 0.10 | 0.58 ± 0.10 | 0.08 ± 0.07 |
| V2 | 3 | MLE | WM | 1302 | 0.18 ± 0.10 | 0.62 ± 0.10 | 0.09 ± 0.07 |
| V2 RS | 3 | MLE | WM | 1337 | 0.18 ± 0.10 | 0.61 ± 0.09 | 0.09 ± 0.07 |
| V1 | 2 | $NN_{MLE}$ | WM | 1022 | 0.21 ± 0.09 | 0.61 ± 0.11 | 0.18 ± 0.09 |
| V2 | 2 | $NN_{MLE}$ | WM | 1289 | 0.21 ± 0.09 | 0.62 ± 0.10 | 0.17 ± 0.07 |
| V2 RS | 2 | $NN_{MLE}$ | WM | 1285 | 0.21 ± 0.09 | 0.64 ± 0.10 | 0.18 ± 0.07 |
| V1 | 3 | $NN_{MLE}$ | WM | 1022 | 0.21 ± 0.09 | 0.59 ± 0.09 | 0.14 ± 0.07 |
| V2 | 3 | $NN_{MLE}$ | WM | 1302 | 0.20 ± 0.09 | 0.64 ± 0.10 | 0.15 ± 0.07 |
| V2 RS | 3 | $NN_{MLE}$ | WM | 1337 | 0.20 ± 0.09 | 0.63 ± 0.09 | 0.15 ± 0.06 |
| V1 | 2 | $NN_{GT}$ | WM | 1022 | 0.22 ± 0.10 | 0.63 ± 0.11 | 0.18 ± 0.07 |
| V2 | 2 | $NN_{GT}$ | WM | 1289 | 0.21 ± 0.10 | 0.63 ± 0.10 | 0.17 ± 0.06 |
| V2 RS | 2 | $NN_{GT}$ | WM | 1285 | 0.21 ± 0.10 | 0.65 ± 0.10 | 0.17 ± 0.05 |
| V1 | 3 | $NN_{GT}$ | WM | 1022 | 0.20 ± 0.10 | 0.59 ± 0.09 | 0.14 ± 0.05 |
| V2 | 3 | $NN_{GT}$ | WM | 1302 | 0.19 ± 0.09 | 0.62 ± 0.10 | 0.15 ± 0.05 |
| V2 RS | 3 | $NN_{GT}$ | WM | 1337 | 0.20 ± 0.10 | 0.62 ± 0.09 | 0.15 ± 0.05 |
| V1 | 2 | MLE | GM | 251 | 0.44 ± 0.14 | 0.42 ± 0.12 | 0.03 ± 0.11 |
| V2 | 2 | MLE | GM | 272 | 0.45 ± 0.11 | 0.46 ± 0.09 | 0.01 ± 0.03 |
| V2 RS | 2 | MLE | GM | 275 | 0.45 ± 0.11 | 0.46 ± 0.10 | 0.01 ± 0.06 |
| V1 | 3 | MLE | GM | 215 | 0.47 ± 0.13 | 0.44 ± 0.11 | 0.01 ± 0.08 |
| V2 | 3 | MLE | GM | 258 | 0.45 ± 0.11 | 0.47 ± 0.11 | 0.01 ± 0.08 |
| V2 RS | 3 | MLE | GM | 251 | 0.46 ± 0.11 | 0.46 ± 0.10 | 0.01 ± 0.07 |
| V1 | 2 | $NN_{MLE}$ | GM | 251 | 0.45 ± 0.14 | 0.39 ± 0.12 | 0.06 ± 0.11 |
| V2 | 2 | $NN_{MLE}$ | GM | 272 | 0.48 ± 0.11 | 0.40 ± 0.05 | 0.02 ± 0.06 |
| V2 RS | 2 | $NN_{MLE}$ | GM | 275 | 0.48 ± 0.12 | 0.41 ± 0.05 | 0.03 ± 0.07 |
| V1 | 3 | $NN_{MLE}$ | GM | 215 | 0.50 ± 0.13 | 0.38 ± 0.09 | 0.03 ± 0.08 |
| V2 | 3 | $NN_{MLE}$ | GM | 258 | 0.48 ± 0.12 | 0.39 ± 0.07 | 0.02 ± 0.08 |
| V2 RS | 3 | $NN_{MLE}$ | GM | 251 | 0.48 ± 0.12 | 0.38 ± 0.05 | 0.02 ± 0.07 |
| V1 | 2 | $NN_{GT}$ | GM | 251 | 0.54 ± 0.14 | 0.44 ± 0.09 | 0.10 ± 0.11 |
| V2 | 2 | $NN_{GT}$ | GM | 272 | 0.55 ± 0.14 | 0.42 ± 0.05 | 0.06 ± 0.05 |
| V2 RS | 2 | $NN_{GT}$ | GM | 275 | 0.55 ± 0.15 | 0.42 ± 0.05 | 0.06 ± 0.07 |
| V1 | 3 | $NN_{GT}$ | GM | 215 | 0.57 ± 0.15 | 0.41 ± 0.07 | 0.06 ± 0.09 |
| V2 | 3 | $NN_{GT}$ | GM | 258 | 0.53 ± 0.15 | 0.40 ± 0.05 | 0.04 ± 0.08 |
| V2 RS | 3 | $NN_{GT}$ | GM | 251 | 0.55 ± 0.15 | 0.40 ± 0.04 | 0.05 ± 0.07 |
| V1 | 2 | MLE | CSF | 145 | | | 0.97 ± 0.11 |
| V2 | 2 | MLE | CSF | 91 | | | 0.97 ± 0.13 |
| V2 RS | 2 | MLE | CSF | 95 | | | 0.97 ± 0.14 |
| V1 | 3 | MLE | CSF | 145 | | | 0.97 ± 0.10 |
| V2 | 3 | MLE | CSF | 93 | | | 0.97 ± 0.15 |
| V2 RS | 3 | MLE | CSF | 87 | | | 0.98 ± 0.11 |
| V1 | 2 | $NN_{MLE}$ | CSF | 145 | | | 0.95 ± 0.12 |
| V2 | 2 | $NN_{MLE}$ | CSF | 91 | | | 0.96 ± 0.12 |
| V2 RS | 2 | $NN_{MLE}$ | CSF | 95 | | | 0.95 ± 0.13 |
| V1 | 3 | $NN_{MLE}$ | CSF | 145 | | | 0.92 ± 0.11 |
| V2 | 3 | $NN_{MLE}$ | CSF | 93 | | | 0.93 ± 0.16 |
| V2 RS | 3 | $NN_{MLE}$ | CSF | 87 | | | 0.91 ± 0.16 |
| V1 | 2 | $NN_{GT}$ | CSF | 145 | | | 0.88 ± 0.08 |
| V2 | 2 | $NN_{GT}$ | CSF | 91 | | | 0.89 ± 0.11 |
| V2 RS | 2 | $NN_{GT}$ | CSF | 95 | | | 0.89 ± 0.10 |
| V1 | 3 | $NN_{GT}$ | CSF | 145 | | | 0.85 ± 0.08 |
| V2 | 3 | $NN_{GT}$ | CSF | 93 | | | 0.85 ± 0.14 |
| V2 RS | 3 | $NN_{GT}$ | CSF | 87 | | | 0.88 ± 0.09 |

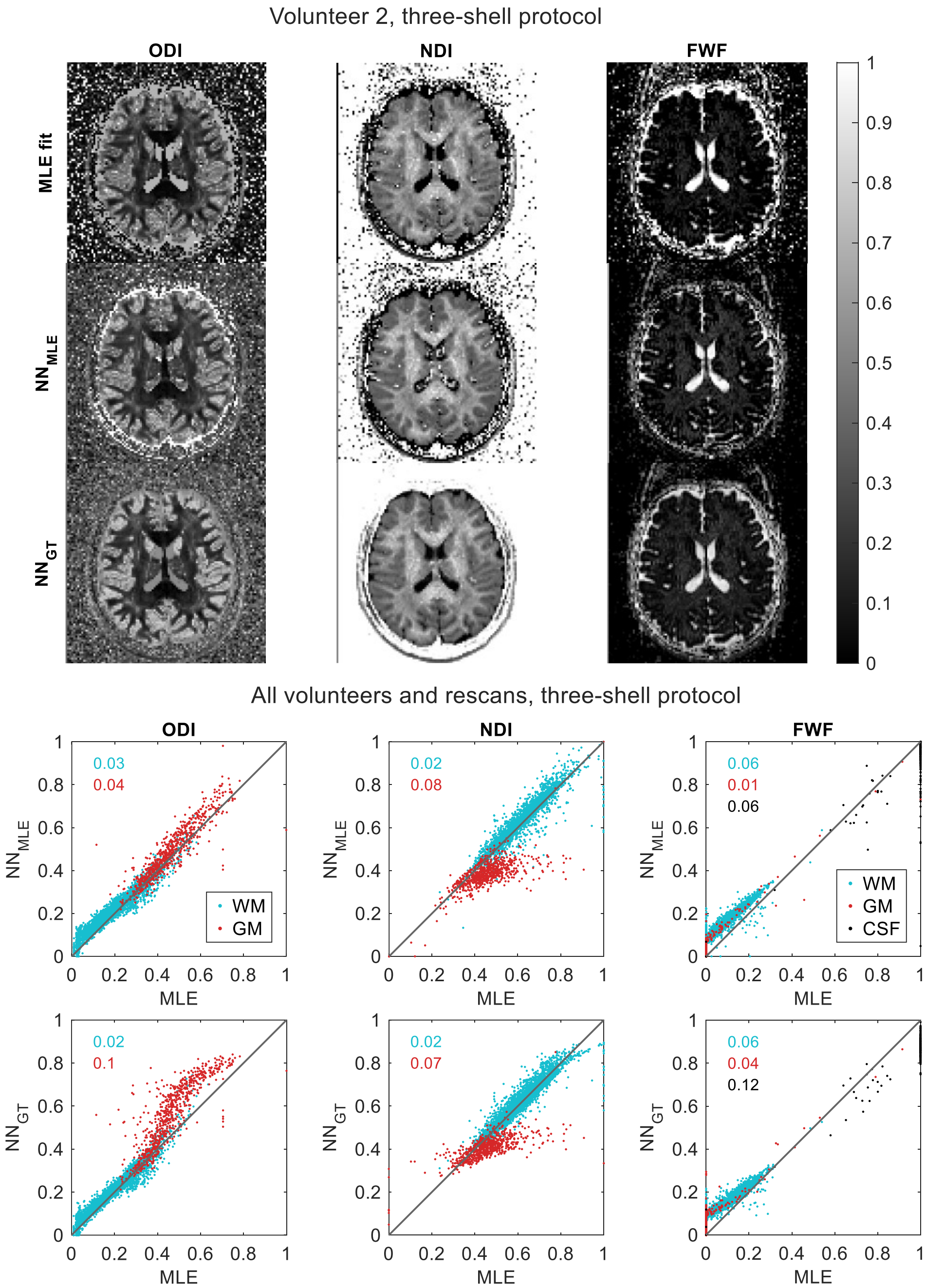

## Figure S1

Estimated orientation dispersion index (ODI), neurite density index (NDI) and free water fraction (FWF) parameter maps for an axial slice of V2 and the three-shell protocol. The upper row shows maps fitted conventionally with the NODDI MATLAB Toolbox (indicated MLE); the next two rows show maps inferred with the two trained neural networks ($NN_{MLE}$ and $NN_{GT}$). The bottom two rows show scatter plots of estimated parameters for each tissue type of single axial slices of all volunteers and rescans using the two-shell protocol, comparing MLE to $NN_{MLE}$ (row 4) and $NN_{GT}$ (row 5). Pairwise MLE-NN mean absolute differences for each tissue class are noted in the top left of each panel. NN parameter maps were exported as DICOMs, converted to NIFTI format and rescaled from integer (range 0 to 1000) to float (range 0 to 1). No post-processing was performed.

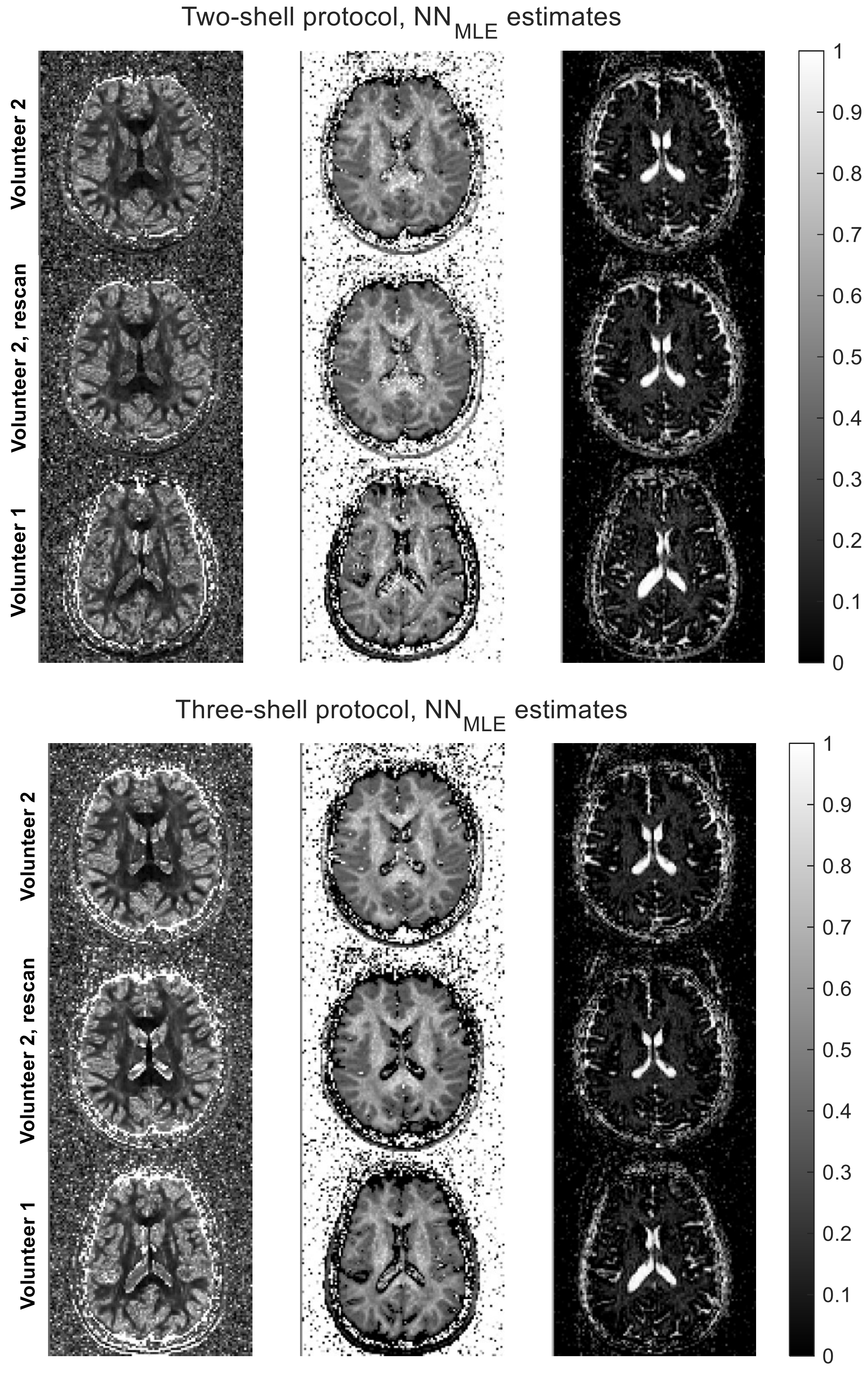

## Figure S2

Estimated orientation dispersion index (ODI), neurite density index (NDI) and free water fraction (FWF) parameter maps for an axial slice of all volunteers, rescans and protocols, computed with the neural network trained on conventionally estimated training labels ($NN_{MLE}$).

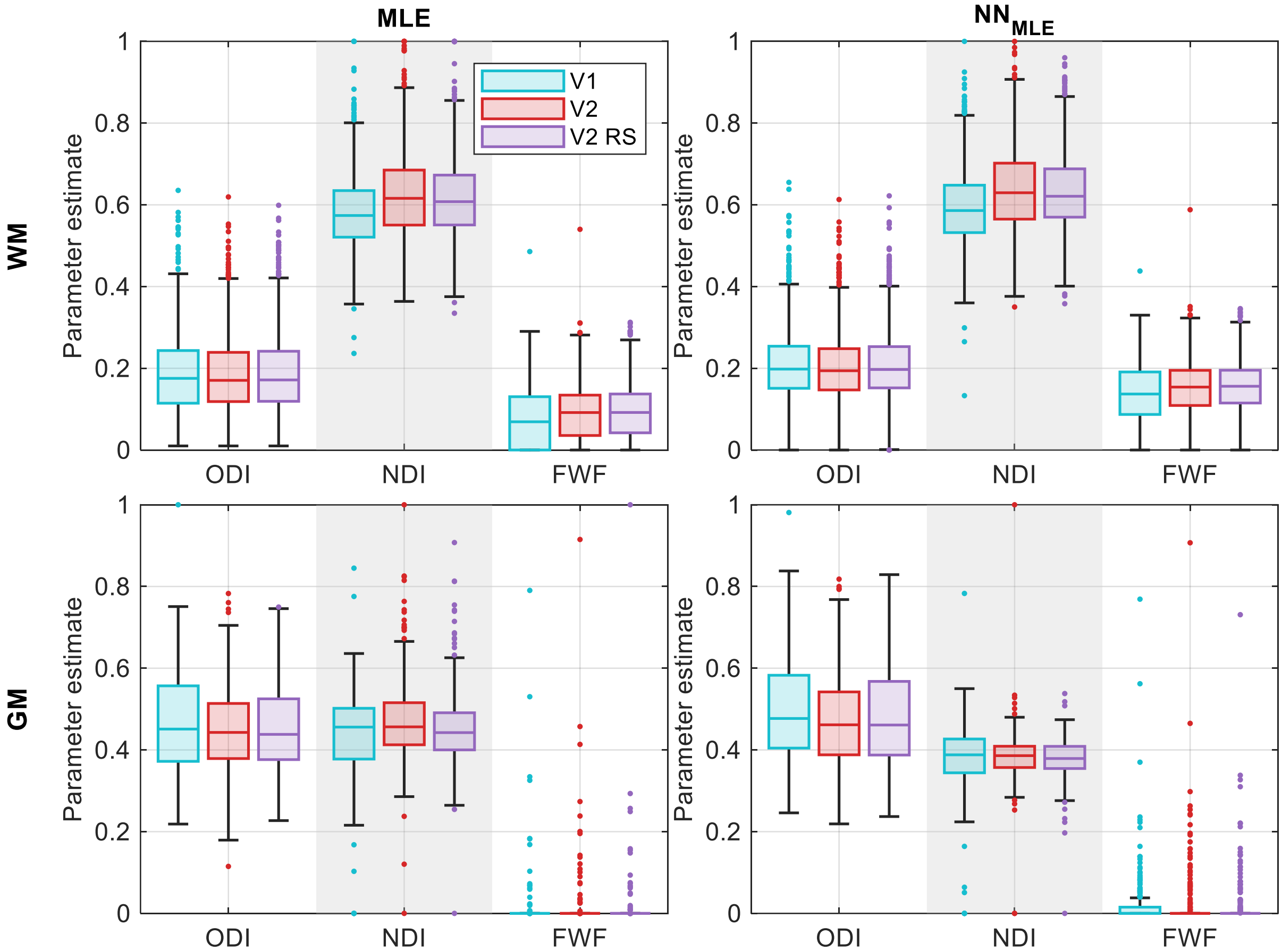

## Figure S3

Boxplots of estimated orientation dispersion index (ODI), neurite density index (NDI) and free water fraction (FWF) parameters for an axial slice across all volunteers and rescans (different colours) of the three-shell protocol. Each panel shows data for a particular tissue type and estimation method, as indicated in row and column labels. The solid horizontal line indicates the median, the box indicates the interquartile range (IQR) and whiskers indicate outlier bounds, 1.5 IQRs from the lower and upper quartiles.

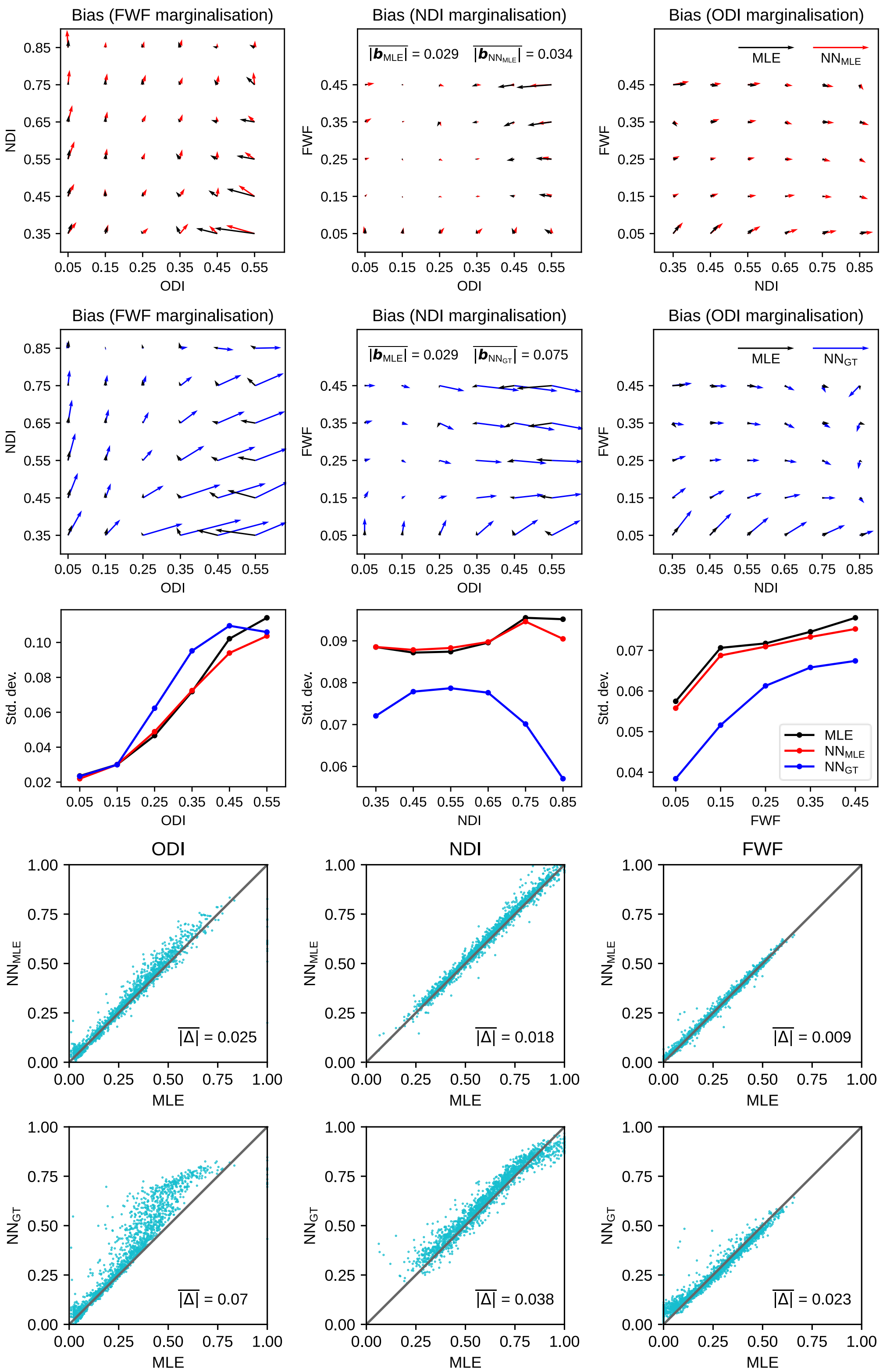

## Figure S4

Results from offline model evaluation for the two-shell protocol. The first two rows show 2D vector plots of the bias (relative to ground truths) in estimated parameters of synthetic test signals for the $NN_{MLE}$ (red) and $NN_{GT}$ (blue) networks, compared to a conventional maximum likelihood estimator (MLE) fit (black). Mean magnitude biases (3D) are displayed in central plots. The middle row shows standard deviations of estimated parameters. Marginalisation was applied across parameter space for the visualisations of the bias (1D marginalisation) and standard deviation (2D marginalisation). The bottom two rows show scatter plots comparing NNs with MLE. To improve clarity, only estimates for the first 10 out of 100 repeats per parameter combination are shown. The mean of absolute differences ($\overline{|\Delta|}$) calculated across all 100 repeats is indicated.

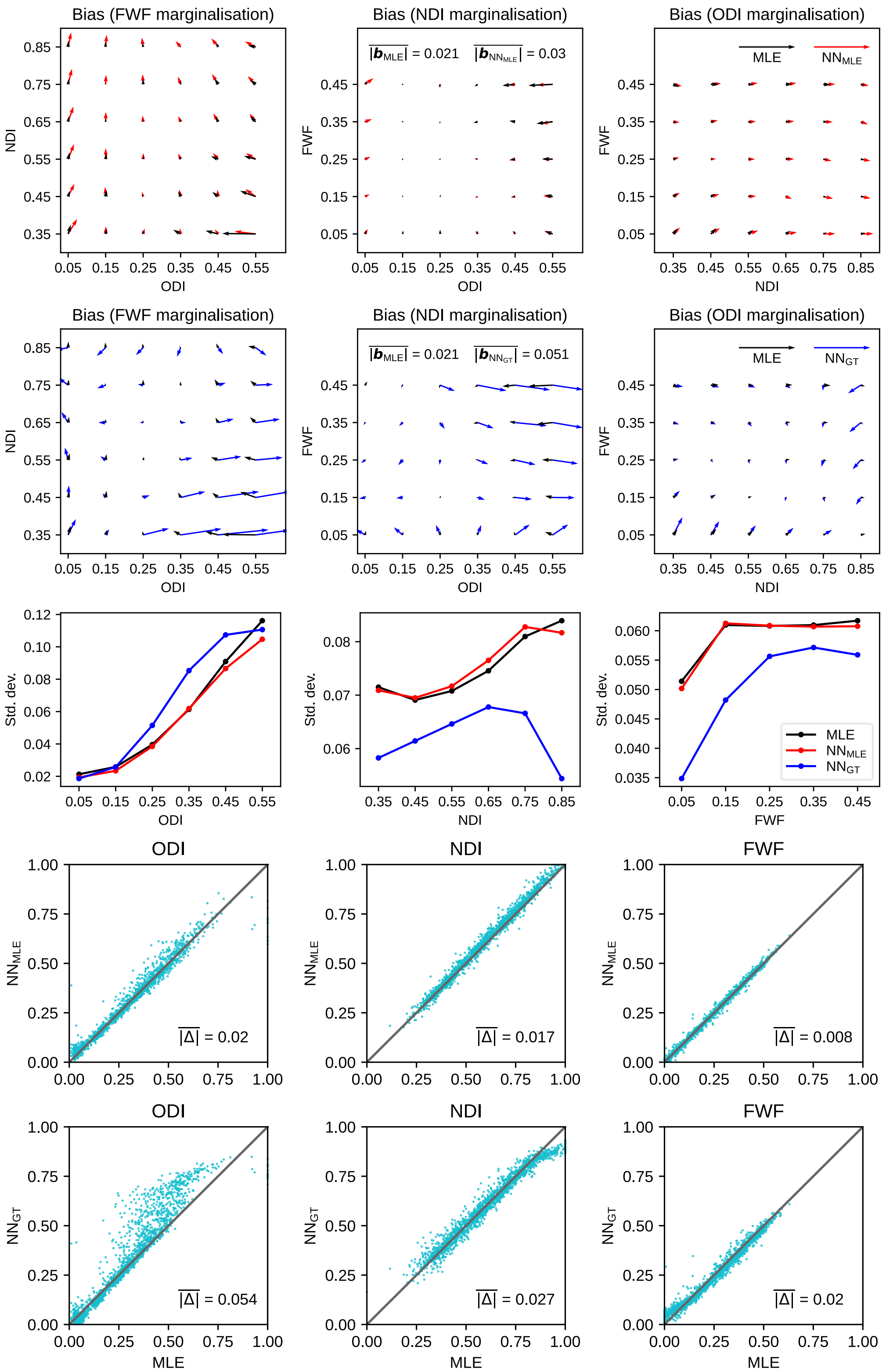

## Figure S5

Brain Results from offline model evaluation for the three-shell protocol. The first two rows show 2D vector plots of the bias (relative to ground truths) in estimated parameters of synthetic test signals for the $NN_{MLE}$ (red) and $NN_{GT}$ (blue) networks, compared to a conventional maximum likelihood estimator (MLE) fit (black). Mean magnitude biases (3D) are displayed in central plots. The middle row shows standard deviations of estimated parameters. Marginalisation was applied across parameter space for the visualisations of the bias (1D marginalisation) and standard deviation (2D marginalisation). The bottom two rows show scatter plots comparing NNs with MLE. To improve clarity, only estimates for the first 10 out of 100 repeats per parameter combination are shown. The mean of absolute differences ($\overline{|\Delta|}$) calculated across all 100 repeats is indicated.